# $F_{max}$ = 270 GHz InAlN/GaN HEMT on Si with forming gas/nitrogen two-step annealing

Peng Cui[1], Meng Jia[2], Guangyang Lin[1], Jie Zhang[1], Lars Gundlach[2], and Yuping Zeng[1*]

[1]Department of Electrical and Computer Engineering, University of Delaware, Newark, DE 19716, USA

[2]Department of Chemistry and Biochemistry, University of Delaware, Newark, DE 19716, USA.

E-mail: yzeng@udel.edu

*Abstract*—In this letter, $N_2$ and forming gas (FG) were used during ohmic contact annealing of InAlN/GaN HEMTs on Si. It is found that $N_2$ annealing offers lower ohmic contact resistance ($R_C$) while FG annealing features lower sheet resistance ($R_{sheet}$). Then FG/$N_2$ two-step annealing was used to achieve a subthreshold swing (*SS*) of 113 mV/dec, an on/off current ($I_{on}/I_{off}$) ratio of ~ $10^6$, a transconductance ($g_m$) peak of 415 mS/mm, a record low drain-inducing barrier lowing (DIBL) of 65 mV/V, and a record high power gain cutoff frequency ($f_{max}$) of 270 GHz on 50-nm InAlN/GaN HEMT on Si.

InAlN/GaN high-electron mobility transistor (HEMTs) on Si substrate have attracted extensive attentions [1-6]. Although Si substrate features a low cost and scaling capability than SiC substrate, the larger lattice mismatch between Si and GaN hindered the epitaxial material quality and device RF performance improvement. A high current gain cutoff frequency $f_T$ of 400 GHz was achieved on a 30-nm InAlN/GaN HEMTs on SiC [7]. A balanced current/power gain cutoff frequency $f_T/f_{max}$ of 348/340 GHz (for E-mode device) and $f_T/f_{max}$ of 302/301 GHz (for D-mode device) were demonstrated on 37-nm InAlN/GaN HEMTs on SiC [8]. To date, $f_T/f_{max}$ of 250/204 GHz [6] and $f_T$ of 310 GHz [2] were demonstrated on InAlN/GaN HEMTs on Si, respectively. This indicates that GaN-on-Si HEMTs technology needs to be drastically improved as compared to the GaN-on-SiC counterpart.

To improve device performance of InAlN/GaN HEMTs on Si, technology of fabrication process and material growth are the two keys. $H_2/N_2$ forming gas (FG) annealing has been widely used in the process manufacturing of GaN HEMTs [9-16]. On one hand, FG can be used for the post-metallization annealing (PMA) to avoid unintentional oxidation, decrease leakage current, and reduce the traps by hydrogen passivation [9-12]. R. Wang *et al*. reported that a reverse gate leakage current of InAlN/GaN HEMT on SiC decreased from $10^{-7}$ to $10^{-12}$ A/mm after PMA and a record high on/off current ($I_{on}/I_{off}$) ratio of $10^{12}$ was achieved [10]. On the other hand, FG annealing can also be used to form ohmic contact [13-16].

In this letter, $N_2$ and forming gas (FG) were used in ohmic contact annealing in InAlN/GaN HEMTs on Si. It is found that $N_2$ annealing offers lower ohmic contact resistance ($R_C$) while FG annealing features lower sheet resistance ($R_{sheet}$). X-ray photoelectron spectra (XPS) showed that FG annealing can remove the surface native oxide, leading to a reduced material sheet resistance. Then a process using FG/$N_2$ two-step annealing was developed, and a reduced subthreshold swing (*SS*), an improved transconductance ($g_m$), a record low drain-inducing barrier lowing (DIBL) of 65 mV/V, as well as a $f_T/f_{max}$ of 125/270 GHz was achieved on 50-nm InAlN/GaN HEMT, resulting in a high $(f_T \times f_{max})^{1/2}$ of 184 GHz among GaN HEMTs on Si.

The epitaxial layer used in this letter was grown by metal organic chemical vapor deposition (MOCVD) on 4-inch Si substrate. It consists of 2-nm GaN cap layer, an 8-nm lattice-matched $In_{0.17}Al_{0.83}N$ barrier layer, a 1-nm AlN interlayer, a 15-nm GaN channel layer, a 4-nm $In_{0.12}Ga_{0.88}N$ back barrier layer, and a 2-μm undoped GaN buffer layer. Device fabrication started with mesa isolation using $Cl_2$-based inductively coupled plasma etching. Ti/Al/Ni/Au stack was deposited and annealed to form alloyed ohmic contacts. The ohmic contact rapid thermal annealing (RTA) process was carried out using Solaris 150 Rapid Thermal Processing System with a ramping speed of 50˚C/s. The system temperature accuracy and stability is ± 2.5˚C and the temperature variation across the entire chamber is ±2.5˚C. Three different types of ohmic contact annealing processes were used on three samples. Sample 1 is annealed at 850˚C for 40s in $N_2$. Sample 2 is annealed at 850˚C for 40s in forming gas (FG: 5% $H_2$ and 95% $N_2$). Sample 3 is annealed at 850˚C first in FG for 20s and then in $N_2$ for 20s. Then these three samples were treated using oxygen plasma treatment. Finally, a Ni/Au T-shaped gate with a gate width ($W_g$) of 2 × 20 μm was fabricated by electron beam lithography. The devices present a source-drain spacing ($L_{sd}$) of 1 μm, a gate-source spacing ($L_{gs}$) of 475 nm, and a gate footprint ($L_g$) of 50 nm, respectively. No passivation process was applied on the reported devices.

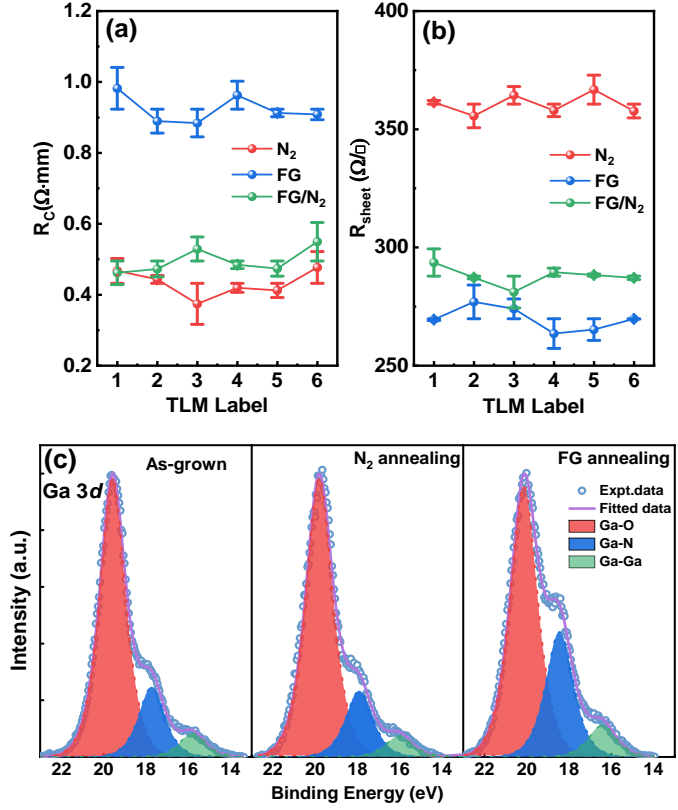

| Table I | $R_C$ ($\Omega\cdot$mm) | $R_{sheet}$ ($\Omega/\square$) | $n_{2D}$ ($cm^{-2}$) | $\mu$ ($cm^2/V\cdot s$) |
|---|---|---|---|---|
| Sample 1 ($N_2$) | 0.43 | 361 | $1.87\times10^{13}$ | 928 |
| Sample 2 (FG) | 0.92 | 270 | $2.15\times10^{13}$ | 1077 |
| Sample 3 (FG/$N_2$) | 0.49 | 288 | $2.11\times10^{13}$ | 1028 |

Fig. 1. (a) Ohmic contact resistance ($R_C$) and (b) sheet resistance ($R_{sheet}$) under $N_2$ annealing, forming gas (FG) annealing, and FG/$N_2$ two-step annealing with six sets of TLM patterns ("I" represents the error bar). (C) XPS spectra of Ga 3$d$ core-level taken from as-grown sample and samples after annealing in $N_2$, and FG, respectively. Table I: the average values of $R_C$ and $R_{sheet}$, electron density ($n_{2D}$), and electron mobility ($\mu$) under different annealing conditions.

The DC current-voltage ($I$–$V$) measurements were carried out by using an Agilent B1500A semiconductor parameter analyzer. Fig. 1(a) and (b) shows the Ohmic contact resistance ($R_C$) and sheet resistance ($R_{sheet}$) for three types annealing conditions obtained using six sets of TLM patterns. The average values of $R_C$ and $R_{sheet}$ (shown in Table I) are 0.43 Ω·mm/361 Ω/□ ($N_2$ annealing), 0.92 Ω·mm/269 Ω/□ (FG annealing), and 0.49 Ω·mm/288 Ω/□ (FG/$N_2$ annealing), respectively. $N_2$ annealing showed the lowest $R_C$ and highest $R_{sheet}$, and FG annealing presented the opposite behavior. Fig. 1(c) compares the X-ray photoelectron spectra (XPS) of Ga 3$d$ core-level taken from as-grown sample (without annealing) and samples after annealing in $N_2$ and FG. The spectra can be resolved into three peaks. The fitting peaks at around 20, 18, and 16 eV binding energies are attributed to Ga-O, Ga-N, and Ga-Ga bonds, respectively [17-19]. The chemical component change of the sample after $N_2$ annealing can be neglected compared to that of the as-grown sample; while for the sample with FG annealing, the Ga-O component significantly decreases and Ga-N component increases, compared to those in the as-grown sample. The results indicate that FG annealing can effectively remove the low quality native oxide (GaO$_x$) and form a nitridation interlayer [19-21]. As shown in Table I, two-dimensional electron gas (2DEG) electron density ($n_{2D}$) extracted from capacitance-voltage measurement increased from 1.87×10$^{13}$ cm$^{-2}$ ($N_2$) to 2.15 × 10$^{13}$ cm$^{-2}$ (FG), presenting a 15% increase. Based on the $R_{sheet}$ and $n_{2D}$, the 2DEG electron mobility ($\mu$) can be calculated and showed a 16% increase with FG annealing, an indication of an improved material surface quality with a weakened remote surface charge scattering [22-25]. However, the increased $R_C$ from FG annealing degrades the device performance. In order to benefit from the good material property due to FG annealing and the low $R_C$ due to $N_2$ annealing, FG/$N_2$ two-step annealing was applied on Sample 3 and a compromised characteristic was obtained.

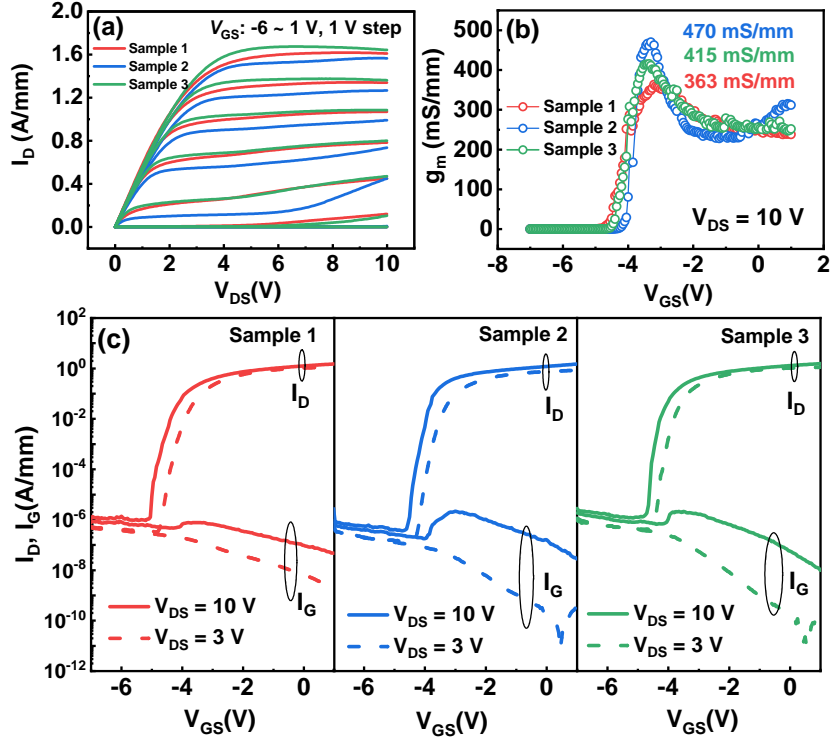

Fig. 2. (a) *I-V* output characteristic; (b) extrinsic transconductance, (c) transfer curves and gate current curves in semi-log scale at $V_{DS}$ = 10 and 3 V of the three samples.

Fig. 2(a) shows the *I-V* output characteristics of the 50-nm InAlN/GaN HEMTs. The device on-resistances ($R_{on}$) extracted at gate-source voltage ($V_{GS}$) of 0 V and drain-source voltage ($V_{DS}$) in the range between 0 and 0.5 V are 1.77 Ω·mm (Sample 1), 2.02 Ω·mm (Sample 2), and 1.82 Ω·mm (Sample 3), respectively. The discrepancy of $R_{on}$ values with those calculated from TLM results should come from the Asher oxidation treatment prior to gate metal deposition, which can oxidize the material surface and then change the channel resistance. The maximum saturation drain currents ($I_D$) at $V_{GS}$ = 1 V and $V_{DS}$ = 10 V are 1.61, 1.57 and 1.64 A/mm for the three samples. The lowest saturation $I_D$ of Sample 2 is due to the high $R_C$. The removed native oxide can effectively increase $n_s$ and gate capacitance, therefore, Sample 3 features the highest saturation $I_D$. Fig. 2(b) shows the extracted extrinsic transconductance ($g_m$) at $V_{DS}$ = 10 V. Compared to $g_m$ peak of Sample 1 (363 mS/mm), Samples 2 and 3 showed higher $g_m$ peak of 470 and 415 mS/mm, respectively, leading to the improved $g_m$ achieved with FG annealing. FG annealing removed the low quality native oxide and reduced gate-to-channel distance, leading to the increased $g_m$ peak. However, the FG/N$_2$ two-step annealing can increase the nitridation layer, which can increase gate-to-channel distance. Therefore, a slightly decreased $g_m$ compared with that of Sample 2 was observed. Fig. 2(c) shows the transfer characteristics and gate current ($I_G$) curves of three samples in semi-log scale at $V_{DS}$ of 10 V and 3 V, respectively. At $V_{DS}$ = 10 V, on/off current ($I_{on}/I_{off}$) ratio of three samples were ~ $10^6$ and an improved subthreshold swing (*SS*) were observed (Samples 1-3: 237, 166, 113 mV/dec, respectively). The drain-inducing barrier lowing (DIBL) of 69, 65, and 65 mV/V were extracted at $I_D$ = 10 mA/mm between $V_{DS}$ = 10 V and $V_{DS}$ = 3V. To the best of our knowledge, this is the lowest value (65 mV/V) among all GaN HEMTs on Si. The breakdown voltage ($BV_{DS}$) of 20 V for three sample was determined at $I_D$ = 1 mA/mm when $V_{GS}$ was fixed at -8 V.

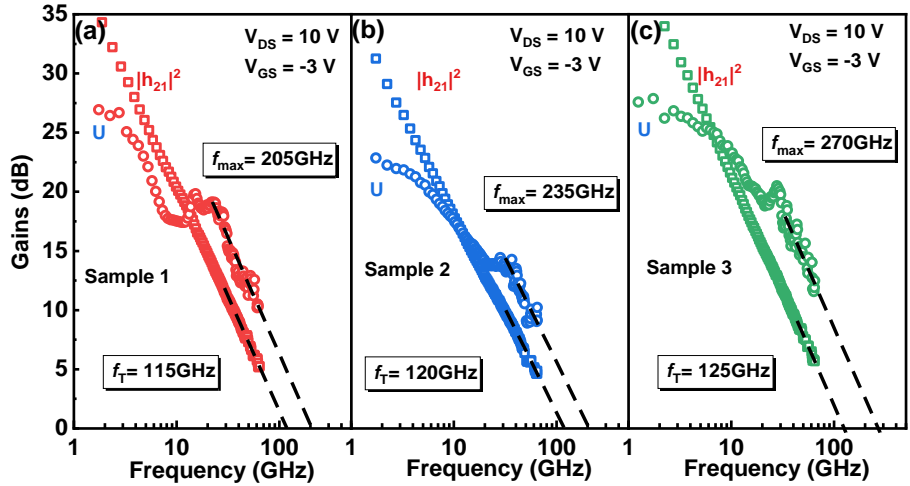

Fig. 3. RF performance of 50-nm InAlN/GaN HEMT at $V_{gs}$ = -3 V and $V_{ds}$ = 10 V with (a) $f_T/f_{max}$ of 115/205 GHz for Sample 1, (b) $f_T/f_{max}$ of 120/210 GHz for Sample 2, and (c) $f_T/f_{max}$ of 125/270 GHz for Sample 3.

The RF measurement was taken with Anritsu MS4647B vector network analyzer configured to operate from 1 to 65 GHz. The network analyzer was calibrated using Line Reflect Match (LRM) calibration. On-wafer open and short structures were used to eliminate the effects of parasitic elements. After de-embedding, the current gain $|h_{21}|^2$ and unilateral gain U as a function of frequency at $V_{DS}$ = 10 V, $V_{GS}$ = −3 V were shown in Fig. 3. $f_T/f_{max}$ of 115/205, 120/210, 125/270 GHz for three samples are extracted using extrapolation of $|h_{21}|^2$ with a -20 dB/dec slope, resulting in $f_T \cdot L_g$ of 5.75, 6, and 6.25 GHz·μm, respectively. Here $(f_T \times f_{max})^{1/2}$ of 154, 159, and 184 GHz were obtained for the three samples. With FG/N$_2$ two-step annealing, $f_T$ increased slightly but $f_{max}$ presented a drastic improvement. Compared with FG annealing, the FG/N$_2$ two-step annealing can facilitate the formation of nitridation interlayer. The nitrogen atoms can effectively decrease the O vacancies from the removal of native oxide and offered better material interface [19], which improves the device performance. To the best of our knowledge, this is the highest $f_{max}$ among reported GaN HEMTs on Si, as shown in Fig. 4.

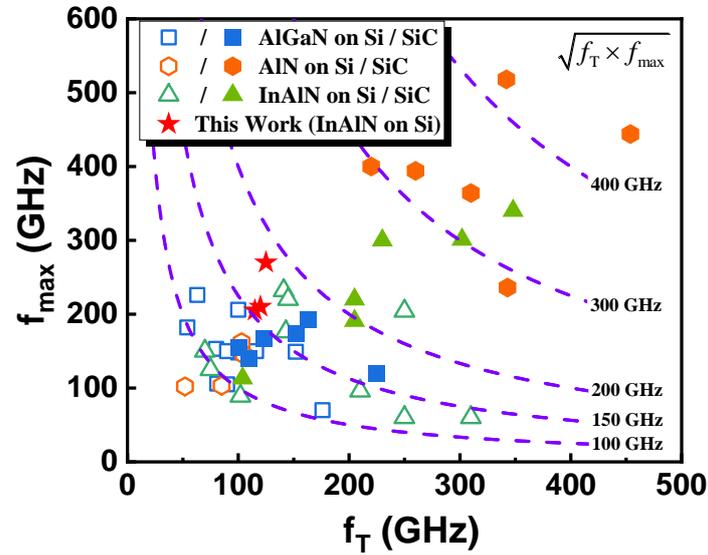

Fig. 4. Comparison of $f_T$/$f_{max}$ of the InAlN/GaN HEMTs on Si in this work with other reported GaN HEMTs (AlGaN on Si [26-36]/SiC [37-40]), AlN on Si [41-43]/SiC [44-48]), and InAlN on Si [2, 4, 49-54]/SiC [55-59])).

In summary, the 50-nm InAlN/GaN HEMT with FG/N$_2$ annealing exhibits an $I_{on}$/$I_{off}$ ratio of $10^6$, a $g_m$ peak of 415 mS/mm, an average SS of 113 mV/dec, and a DIBL of 65 mV/V. RF measurement of the 50-nm InAlN/GaN HEMT presents a $f_T$/$f_{max}$ of 125/270 GHz and an $(f_T \times f_{max})^{1/2}$ of 184 GHz. The fabrication technology for GaN HEMTs on Si yields excellent RF characteristics, which shows the great application potential of GaN-on-Si for millimeter wave power amplifiers.